\documentclass[12pt]{article}
\usepackage{setspace}
\usepackage{amsmath,amssymb,mathrsfs}
\usepackage{color}
\usepackage{hyperref}
\usepackage{graphicx,epsfig}
\numberwithin{equation}{section}
\begin{document}
\setlength{\topmargin}{-1cm} 
\setlength{\oddsidemargin}{-0.25cm}
\setlength{\evensidemargin}{0cm}
\newcommand{\e}{\epsilon}
\newcommand{\beq}{\begin{equation}}
\newcommand{\eeq}[1]{\label{#1}\end{equation}}
\newcommand{\bea}{\begin{eqnarray}}
\newcommand{\eea}[1]{\label{#1}\end{eqnarray}}
\renewcommand{\Im}{{\rm Im}\,}
\renewcommand{\Re}{{\rm Re}\,}
\newcommand{\diag}{{\rm diag} \, }
\newcommand{\Tr}{{\rm Tr}\,}
\def\draftnote#1{{\color{red} #1}}
\def\bldraft#1{{\color{blue} #1}}
\def\n{n \cdot v}
\def\ni{n\cdot v_I}
\begin{titlepage}
\begin{center}

\vskip 4 cm

{\Large \bf Local Operator Algebras of Charged States in Gauge Theory and Gravity}

\vskip 1 cm

{P.A. Grassi$^{a,b,c}$~\footnote{E-mail: \href{mailto:pietro.grassi@uniupo.it}{pietro.grassi@uniupo.it}} and M. Porrati$^d$~\footnote{E-mail: \href{mailto:mp9@nyu.edu}{mp9@nyu.edu}} }

\vskip .75 cm

{ $^{(a)}$ \it Dipartimento di Scienze e Innovazione Tecnologica (DiSIT),\\
Universit\`a del Piemonte Orientale, viale T. Michel, 11, 15121 Alessandria, Italy

 {$^{(b)}$
 \it Theoretical Physics Department, CERN, 1211 Geneva 23, Switzerland}

 {$^{(c)}$
 \it INFN,} 
 {\it Sez. Torino,} 
 {\it via P. Giuria, 1, 10125, Torino, Italy}

$^{(c)}$ 
Center for Cosmology and Particle Physics,  Department of Physics, New York University, \\ 726 Broadway, New York, NY 10003, USA}

\end{center}

\vskip 1.25 cm

\begin{abstract}
\noindent  
Powerful techniques have been developed in quantum field theory that employ algebras of local operators, yet local 
operators cannot create physical charged states in gauge theory or physical nonzero-energy states in perturbative quantum
 gravity. A common method to obtain physical operators out of local ones is to dress the latter using appropriate Wilson lines.
 This procedure destroys locality, it must be done case by case for each charged operator in the algebra, and it rapidly 
 becomes cumbersome, particularly in perturbative quantum gravity. 
 In this paper we present an alternative approach to the definition of physical charged operators: we 
 define an automorphism that maps an algebra of local charged operators into a (non-local)  algebra of physical charged 
 operators. The automorphism is described by a formally unitary intertwiner mapping the exact BRS operator 
 associated to the gauge symmetry into its quadratic part. 
 
 The existence of an automorphism between local operators and the physical ones,
 describing charged states, allows to retain many of the results derived in local operator algebras and extend them to the 
 physical-but-nonlocal algebra of charged operators as we discuss in some simple applications of our construction. We also
 discuss a formal construction of physical states and possible obstructions to it.
  \end{abstract}
\end{titlepage}
\tableofcontents
\newpage

\section{Introduction}\label{intro}
The definition of gauge-invariant physical states in gauge theories has a long history that dates back to the early days of 
quantum field theory. The problem with such states is in fact already clearly seen in the textbook example given
by Dirac~\cite{dirac}, of a charge-$q$ operator $\psi(x,t)$ in QED. To make it gauge invariant we can dress it with the 
classical electromagnetic field created by a point-like charge-$q$ source at position $x$ and
 time $t$. This produces the gauge invariant operator 
\beq
\psi^{inv}(x)=\exp\left[-iq\int d^3y A^i(y,t) (y^i-x^i)/|y-x|^3\right]\psi(x,t).
\eeq{intro1}
The gauge-invariant field is obviously nonlocal. In fact the Gauss law implies that the electric charge is a 2D surface integral
computed at space-like infinity and hence it commutes with all local operators defined on any space-like 3D volume bound
by the surface. Neither dressing with Wilson lines nor re-phrasing the problem in a more modern language in terms of
BRST-invariant operators~\cite{BRS} can change this fact.

In gauge theories local and semilocal operators carrying zero charge define a rich algebra
whose representations define superselection sectors of the theory Hilbert space. This may be enough to fully define QED in an algebraic setting~\cite{buch}. On the other hand, nonlocal operators are necessary to
connect different superselection sectors and in fact to make precise the very definition of charge. These operators are ubiquitous and fundamental in gauge theories but the need to give up locality to define charged 
states has some drawbacks. This is particularly true in quantum gravity, where energy itself can be written as a 2D surface 
integral at space-like infinity~\cite{dirac2,ADM1,ADM2} so that {\em all} nonzero energy states are necessarily nonlocal.

The dressing necessary
to make any specific local operator commute with the BRST charge associated to general coordinate transformations can 
be defined in principle as a perturbative series in the Newton constant around a fixed background 
(see e.g.~\cite{donn,giddings}) but the construction becomes rapidly unwieldy, it converges only as a formal power 
series, and it has to be done anew for each local operator. 

\begin{figure}[h]
\begin{center}
\epsfig{file=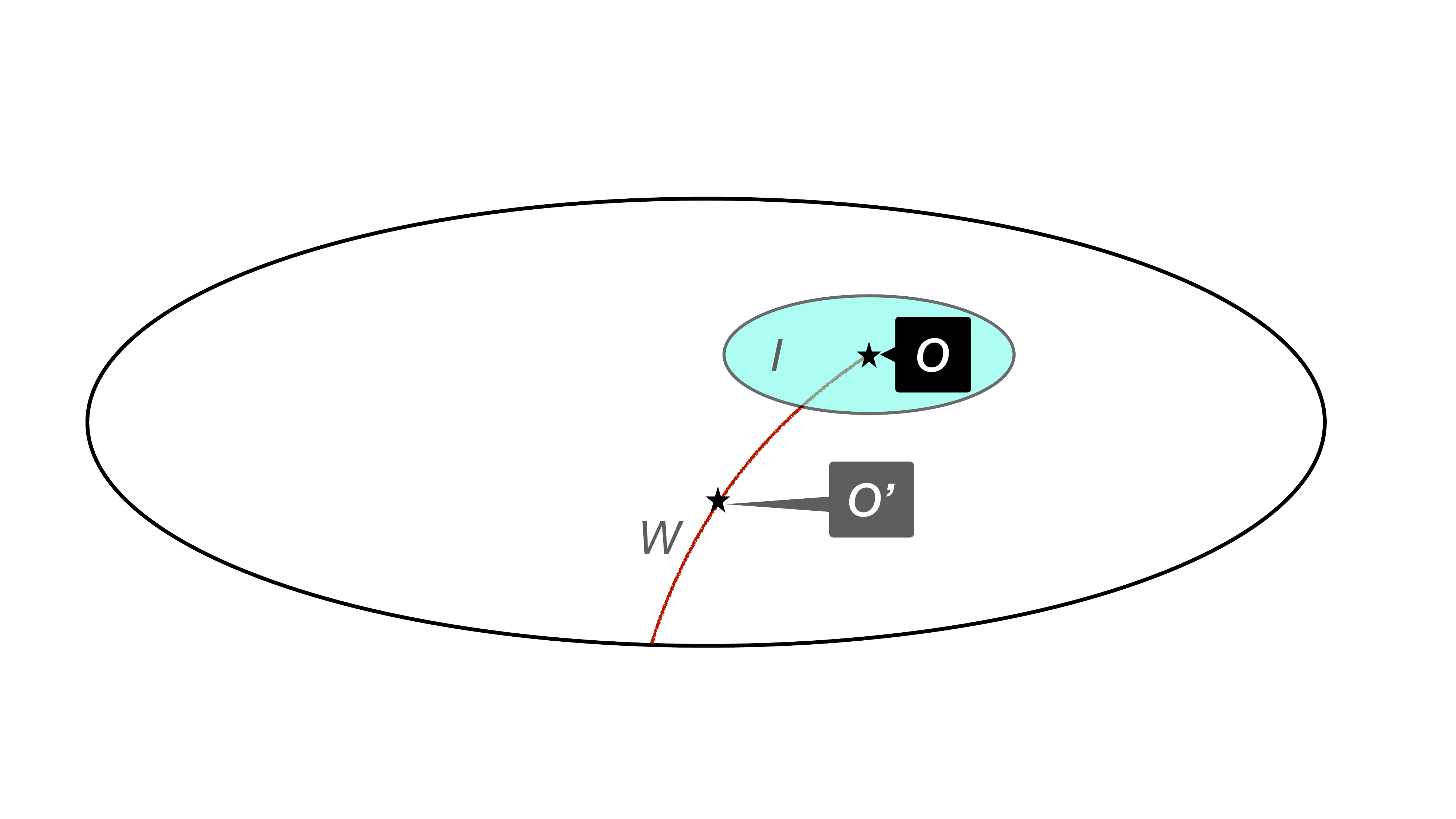, height=3in, width=5in}
\end{center}
\caption{The dressing $W$ of a local operator $O$ inside an island $I$ must ``jump out'' of it to reach space-like infinity, so 
it may cross some operators $O'$ living outside the island.}
\label{island}
\end{figure}

The dressing of charged states requires either an infinite Wilson line or a dressing by a classical field extending to infinity.
So it  cannot be used to define states in a bounded subregion of a Cauchy surface (an ``island'') commuting with {\em all} physical 
operators  outside the region: some of the outside operators are going to ``intercept'' the Wilson line, see figure~\ref{island}, 
and generically not commute with it. On the other hand, if the states outside the island are to be independent of those inside,
they should be mutually commuting. So, this manifestation of the breakdown of locality inherent in the construction of physical
states appears to be in contradiction with the assumption that in semiclassical gravity states localized inside the island are 
independent of those localized outside it. Commutativity of space-like separated operators is used to argue that the bulk 
entropy is extensive and the latter
property is necessary to define the quantum extremal surfaces used to recover the Page unitarity 
bound~\cite{page} in black-hole evaporation~\cite{alm}. This problem (and potential solutions) was noticed in~\cite{randall}.

Another problem with nonlocal operators is that many fundamental results obtained in the algebraic approach to quantum 
field theory use locality in an essential way. When locality is lost it becomes unclear if all or some or any of the properties
usually associated with the algebra of local operators in relativistic quantum field theory or perturbative quantum gravity 
survive. 

In this paper we address the problems due to nonlocality of charged states in a novel way by introducing a new definition of gauge invariant operators. Instead of dressing each operator on a 
case-by-case basis we define an automorphism of algebras that maps local operators into gauge-invariant ones. 
The automorphism is defined by a formally unitary operator $\Omega$ acting on operators in the canonical fashion
\beq
O\rightarrow O'=\Omega^\dagger O \Omega, \quad \Omega^\dagger\Omega=1, \quad O\in \mathscr{A} .
\eeq{intro2}
Since the map $O\rightarrow O'$ is an automorphism of algebras, it preserves the properties of the local operator algebra
$\mathscr{A}$. So in particular it preserves its von Neumann type and the set of affiliated operators. 

The local algebra $\mathscr{A}$ is {\em by construction} the von Neumann algebra of all local operators that commute with a simplified 
quadratic BRST charge, $Q_0$. In a gauge theory this charge acts linearly, so it only transforms the gauge fields as 
$[Q_0, A_\mu]= \partial_\mu c$. It leaves all other fields invariant; in particular, it leaves a charged field $\psi$ invariant
because its BRST transformation law is quadratic in the fields ($[Q^B,\psi]_{\pm}=c\psi$). The operator $\Omega$ is then defined
as an intertwiner between $Q_0$ and the exact BRST charge $Q^B$:
\beq
Q_0\Omega=\Omega Q^B  .
\eeq{intro3}
The exact BRST charge acts nontrivially on all charged matter operators; explicitly, for an operator $O^q(x)$ of 
charge $q$ it is $Q_{BRST}O^q(x) = qc(x)O^q(x)$. In this paper we choose $Q_0$ to be the BRST charge of a free theory, i.e. the vanishing gauge charge limit for gauge theories or the vanishing Newton constant limit at finite energy for gravity. More
general choices will be discussed elsewhere.
In QED $Q^B$ is a power series in the electric charge but in general we find it useful to construct $Q^B$ more abstractly as a sum of terms $Q^B_n$, $n>0$  with charge $n$ under an appropriately defined charge operator, which we call $S$ and define in subsection~\ref{conj}. In this paper $Q^B$ will always denote the exact BRST charge of the interacting theory.

The map $\mathscr{A} \rightarrow \mathscr{A}'=\Omega^\dagger \mathscr{A} \Omega$
also preserves the defining property of locality: two operators $O_1(x_1)O_2(x_2)\in \mathscr{A}$ 
defined at space-time separated points $x_1,x_2$ commute $[O(x_1),O_2(x_2)]=0$ --or anticommute if both are fermions. 
Since the map is an automorphism, the same is obviously true for $\Omega^\dagger O_1(x_1)\Omega$, 
$\Omega^\dagger O_2(x_2)\Omega$.

The existence of an intertwiner can be argued on a formal basis as follows. The operator $Q_0$ is the BRST charge acting 
on asymptotic states; its linearity is the key to prove positivity of the Hilbert space of in- and out- physical 
states~\cite{BRS,ko}. The formal theory of scattering defines in- and out M\o{}ller scattering operators which intertwine 
between asymptotic operators and finite-time ones. In particular either the in- or out scattering operator precisely 
obeys~\eqref{intro3}. The argument we gave is at best an encouragement because the very existence of scattering 
operators is beset by well known problems~\cite{haag,rs} and it is also an overkill, 
because it also intertwines between a free
Hamiltonian, $H_0$, and an interacting one, $H$. This requires $H$ to be isospectral with $H_0$, a condition that is already
 violated in perturbation theory when $H$ has bound states. 
 
 In this paper we do not look for a scattering operator; instead, we construct an operator that only intertwines between 
 $Q_0$ and $Q^B$. We will work in canonical quantization so all our operators will be defined on a space-like surface 
 $\Sigma$, which
 we take to be the maximal Cauchy surface of the background spacetime. In relativistic quantum field theory a smearing 
 over time is necessary (and sufficient~\cite{borch,witt-reg}) to define operators acting on a dense subset of the Hilbert 
 space so the operators we call $O(x,t)$, $(x,t)\in \Sigma$ are really a shortcut for 
 \beq
 O(x,t)=\int ds f(t-s) O(x,s)=\int ds f(t-s)e^{iH(s-t)} O(x,t) e^{-iH(t-s)}, 
 \eeq{intro4}
with $f(s)$ a smooth, compact support function.\footnote{It is possible that a covariant generalization of the equal-time 
formalism described in this paper may be achieved by employing the techniques described in~\cite{buch}, and could be 
necessary to make our construction rigorous.}

This paper begins with section~\ref{alg}, which itself begins with subsection~\ref{def}, where a review of the construction of the free and interacting BRST charges for
an Abelian $U(1)$ theory minimally coupled to charged matter in the 
Hamiltonian operatorial formalism is given. The next subsection,~\ref{conj}, introduces one of the key 
quantities needed for our construction, namely $R$, an operator of ghost number $-1$ whose anticommutator with the free
charge $Q_0$ is a simple counting operator, $S$, under which the exact charge $Q^B$ decomposes into a finite sum of 
terms
of positive or zero charge. Subsection~\ref{cauchy} extends the construction done in the previous subsections using the surface
$t=0$ in Minkowski space to a Cauchy surface in a generic spacetime $\mathcal{M}$ admitting a foliation 
$\mathcal{M}=[0,1]\times \mathcal{M}_3$ (or $\mathcal{M}=\mathbb{R}\times \mathcal{M}_3$).

Section~\ref{nonab} extends the construction of $Q_0$, $Q^B$, $R$ and $S$ to non-Abelian gauge theories while the next,
\ref{pqg} does the same for perturbative quantum gravity. Specifically, Subsection~\ref{mink} constructs the charges perturbatively around a Minkowski space background while~\ref{maxsym} repeats the construction for $AdS_4$ and outlines
a similar construction for $dS_4$. We leave the case of a generic maximal Cauchy surface for future study. 

Section~\ref{inter}, details in subsection~\ref{int1} the explicit construction of the
intertwiner $\Omega$. In~\ref{int2} it is shown how to use the intertwiner to count physical states 
inside  an island and to compute an index associated to local gauge invariant operators. The final section~\ref{other}
describes, in~\ref{otherdef}, a different procedure for defining local gauge invariant operators that has been adopted either explicitly or 
implicitly in some recent literature. {Subsection~\ref{closed} elaborates on the additional constraints which are 
present when the maximal Cauchy surface of spacetime is a closed manifold.}
Subsection~\ref{states} outlines the steps 
necessary to define physical states in the cohomology of the BRST operator $Q^B$ using the intertwiner $\Omega$, as well
as possible obstructions to it.
Appendix~\ref{app} contains a general overview of the BRST techniques used in our paper.

\section{BRST Charge Algebra for QED}\label{alg}

\subsection{BRST charges as operators on Hilbert spaces}\label{def}

We introduce a generic gauge fixing by modifying the action of QED as follows 
\begin{eqnarray}
\label{EMCA0}
\mathcal L = - \frac14 F_{\mu\nu} F^{\mu\nu} + \rho \partial_\mu A^\mu + \frac\xi2 \rho^2 - 
b \partial_\mu \partial^\mu c { -\bar \psi \gamma^\mu \nabla_\mu \psi }
\end{eqnarray}
leading to the equations 
\begin{eqnarray}
\label{EMCA}
&& \partial^\mu (\partial_\mu A_\nu - \partial_\nu A_\mu) 
- \partial_\nu \rho  - i e \bar\psi \gamma_\nu \psi =0\,, \nonumber \\
&& { \xi \rho + \partial_\mu A^\mu =0}\,,  \nonumber \\
&&  \partial^\mu \partial_\mu c =0\,,  \\
&&  \partial^\mu \partial_\mu b =0\,,  \nonumber \\
&&\gamma^\mu \nabla_\mu \psi=\gamma^\mu (\partial_
\mu + i e A_\mu) \psi = 0\, . \nonumber 
\end{eqnarray}
We introduced the Nakanishi-Lautrup auxiliary field $\rho$ and the antighost $b$. The constant 
$\xi$ parametrizes the choices of the gauge fixing. The action is no longer invariant under the gauge symmetry, but it
possesses instead the BRST symmetry
\begin{eqnarray}
\label{EMCAB}
s A_\mu = \partial_\mu c\,, ~~~~~~
s c= 0\,, ~~~~~~
s b = \rho\,, ~~~~~
s \rho = 0\,,~~~~
s \psi = i e c \psi\,. 
\end{eqnarray}
which is clearly nilpotent since $c^2 =0$. 

For the operator formalism, we need to compute the BRST current. This can be easily done by introducing a 
local anticommuting parameter $\Lambda(x)$ and extracting from the action terms which are proportional to its derivatives.
We find  
\begin{eqnarray}
\label{EMCAC}
j^{_{BRST}}_\mu = \partial^\nu c F_{\nu\mu} + \rho \partial_\mu c - i e \bar\psi \gamma_\mu \psi {c} ,
\end{eqnarray}
which is conserved on-shell using eqs. \eqref{EMCA}. 

It is convenient to consider the auxiliary field $\rho$ as an independent degree of freedom. 
Since the Lagrangian density is 
\beq
\mathcal L = +\rho \dot{A}^0 + \dot{b} \dot{c} + \mbox{terms without time derivatives}
\eeq{mass1}
$\rho$ is the canonically conjugate momentum of $A^0$, $E^i=F^{0i}=-\dot A^i - \partial^i A^0$ is canonically conjugate to $A_i$,  $\dot{b}$ is canonically conjugate to $c$ and $\dot{c}$ is conjugate to $b$ so we get the equal-time (anti)commutators
\bea
[\rho(x),A^0(y)] &=& -i \delta^3( x-y) , \quad [E^i(x),A_j(y)]=-i\delta^i_j \delta^3(x-y) 
\nonumber \\
\quad [\dot{c}(x), b(x)]_+ &=& \delta^3(x-y),\quad [\dot{b}(x), c(y)]_+=-\delta^3(x-y).
\eea{mass2}

Next, let's split the BRST charge $Q^B=\int d^3x j_{BRST}^0=Q_0+Q_1$ into the quadratic part $Q_0$ and all the terms $Q_1$ that are cubic or higher order in the fields. Explicitly
 \beq
 Q_0 = \int d^3x( c\partial_i E^i -\rho \dot{c} ) , \qquad Q_1 = -e \int d^3x c \psi^\dagger \psi .
 \eeq{mass3} 

\subsection{The conjugate operator}\label{conj}

We look for an operator $R$ such that $[R,Q_0]_+\propto S$ with $S$ a bosonic operator with simple commutation 
properties with $Q_1$.
The operator can contain an inverse Laplacian because it does not have on-shell poles, which are instead present in the 
Green functions of wave operators. Our ansatz is
\beq
R=\int d^3x \Big[b(x) A^0(x)  + \dot{b}(x)\int d^3 z \beta(x-z) \partial_i A^i(z) \Big] ,
\eeq{mass5}
with a yet to be defined function $\beta$.

Next we write $[Q_0,R]_+=A+B$ with
\bea
A &=& \int d^3x d^3y d^3z \Big[ \dot{b}(y) \beta(y-z) \partial_i A^i(z) , c(x) \partial_j E^j(x) \Big]_+ \nonumber \\
B &=& \int d^3x d^3y \Big[ b(y)A^0(y) , -\rho(x) \dot{c}(x) \Big]_+  .
\eea{mass6}
Using the (anti)commutation relations in~\eqref{mass2} we get
\bea
A &=& A_1+A_2, \nonumber \\
A_1 &=&- \int d^3 x d^3 y d^3 z  \delta^3(x-y) \beta(y-z) \partial_j E^j(x) \partial_iA^i(z) =
-\int d^3x d^3z  \partial_j E^j(x) \beta(x-z)  \partial_iA^i(z), \nonumber \\
A_2 &=& i\int d^3 x d^3 y d^3 z \dot{b}(y) c(x)\Big[{\partial \over \partial x^j} {\partial \over \partial z_j}\delta(z-x) \Big]\beta(y-z)   =-i \int d^3x d^3y \dot{b}(y) c(x) \Delta_x \beta(y-x) ,\nonumber \\ && 
\eea{mass8}
where we defined $\Delta=\partial_i \partial^i$.
 Term $B$ is
\bea
B &=&  -\int d^3x d^3y \Big[ b(y)A^0(y) \rho(x) \dot{c}(x) +  \rho(x) \dot{c}(x) b(y)A^0(y) \Big] \nonumber \\ &=&
-\int d^3x \rho(x) A^0(x) + \int d^3x d^3y b(y) \dot{c}(x) [\rho(x),A^0(y)]  \nonumber \\ 
&=& -\int d^3x \rho(x) A^0(x) -i \int d^3x b(x) \dot{c}(x) .
\eea{mass10}
If we set $\Delta_x\beta(x-y) =-\delta^3(x-y)$  
we arrive at our final formula
\beq
[Q_0,R]_+=A_1+A_2+B = i \int d^3x \Big[  \dot{b}(x) c(x) -b(x) \dot{c}(x) +i \rho(x)A^0(x) +i \Pi(x) \Phi(x)\Big]\equiv i S.
\eeq{mass11}
Here $\Pi(x) \equiv \partial_i E^i (x)$, $\Phi(x)\equiv\int d^3z \beta(x-z) \partial_i A^i(z)$; they are canonically conjugate
\beq
[\Pi(x),\Phi(y)]=-i\delta^3(x-y).
\eeq{mass11a}
The operator $S$ is an anti-Hermitian number counting operator that counts each ghost $c,\dot{c} $ with a $+1$, each ghost $b,\dot{b}$ with a $-1$, each
$A^0$ and $\Phi$ with a $+1$ and each $\rho$ and $\Pi$ with a $-1$.  
As a simple check of our algebra we see that it commutes, as it obviously should, with $Q_0$, {while $Q_1$ has
charge $+1$: $[S,Q_1]=Q_1$.}

\subsection{Generic Cauchy surfaces and covariant formulas for the BRST-algebra operators}\label{cauchy}

In this subsection, we assume a foliation of the spacetime $[0,1] \times \mathcal{M}_3$, where $ \mathcal{M}_3$ is a 
three-dimensional manifold with induced metric $g = g_{ij} dx^i dx^j$ (Euclidean signature)
{and the spacetime metric is $ds^2= dt^2 + g$}.
We define the Hodge dual operator on a $p$-form as follows 
\begin{eqnarray}
\label{HA}
\star \omega^{p} = \omega_{i_1 \dots i_p} \frac{\sqrt{{\rm det} (g_{ij})}}{(3-p)!}
g^{i_1 j_1} \dots g^{i_p j_p} \epsilon_{j_1 \dots j_3}   dx^{j_{p+1}} \wedge \dots \wedge dx^{j_{3}} ,
\end{eqnarray}
 which is a $3-p$ form. 
 Given this definition, we can also define the conjugate $d^\dagger =(-1)^p \star d \star$ acting on $p$-forms 
 and the Laplace-Beltrami differential 
 $-\Delta = d^\dagger d + d d^\dagger$.  We also define the $3$-form volume ${\rm vol}_3 = \star 1$ and the volume 
 ${\rm Vol}({\mathcal M}_3) = \int_{{\mathcal M}_3} \star 1 =  \int_{{\mathcal M}_3} {\rm vol}_3   = \int \sqrt{g} d^3x$. 
  
 Let us first consider the gauge fixing Lagrangian where $\rho$ is the Nakanishi-Lautrup field and $c,b$ are the ghost and 
 antighost, respectively (see also \cite{KIP}). 
\begin{eqnarray}
\label{HF}
{\mathcal L} &=& 
- \frac12 (dA^0 - \partial_t A)\wedge \star (dA^0 - \partial_t A)+ 
\frac12 d A\wedge \star d A + { \frac \xi 2 \rho \star \rho} \nonumber \\ 
&+& \star \rho ( d^\dagger A + \partial_t A^0) 
-\star b \Big(\partial^2_t c - \Delta c \Big) - A \wedge \star j - A^0 \wedge \star j^0  .
\end{eqnarray}
{ Choosing for simplicity $\xi=0$,} the first line reduces to the conventional Maxwell Lagrangian decomposed into the 
scalar potential $A^0$ and 
the vector potential $A$.
The Maxwell and ghost equations are
\begin{eqnarray}
\label{HG'}
&&d \star (d A^0 - \partial_t A) - \star  \partial_t \rho = \star j^0 , \nonumber \\
&& - \partial_t \star (d A^0 - \partial_t A) + d \star d A + \star d \rho = \star j , \nonumber \\
&& d^\dagger A + \partial_t A^0 =0 , \nonumber \\
&& \partial_t^2 c - \Delta c = 0 ,\nonumber \\
&& \partial_t^2 b - \Delta b =0 .
\end{eqnarray}

Using the gauge fixing (third equation), we can 
rewrite the first two equations as 
\begin{eqnarray}
\label{HI}
&&\partial^2_t A^0 - \Delta A^0 - \partial_t \rho = j^0 , \nonumber \\
 &&\partial^2_t A - \Delta A - d\rho = j ,
\end{eqnarray}
which are conventional hyperbolic wave equations. 
Introducing the $1$-form (gauge invariant) conjugate momenta 
$E =(d A^0 -  \partial_t A)$ we can finally define the BRST 
charge {
 \bea
Q^B &=&  Q_0+Q_1 , \nonumber \\ Q_0 &=& \int_{{\mathcal M}_3} \Big( c \wedge d \star E - \star \rho \partial_t c \Big)  = 
\int_{{\mathcal M}_3} 
\left( \partial_i c g^{ij}(\partial_j A^0 - \partial_t A_i) - \rho \partial_t c\right) \sqrt{g} d^3x , \nonumber \\ 
Q_1 &=& \int_{{\mathcal M}_3} \star j  c =  \int_{{\mathcal M}_3} \sqrt{g} j^0 c  .
\eea{HB}
}
Using the commutation relations 
\begin{eqnarray}
\label{HE}
&&
\Big[ \star \rho(t, \vec{x}), A^0(t, \vec{y}) \Big]=  - { \frac i {\sqrt{g}}} \delta^3(x-y) \star 1\,,  \hspace{1cm} 
\Big[ \star E(t, \vec{x}), A(t, \vec{y}) \Big] =  - {\frac  i {\sqrt{g}}} \delta^3(x-y) \star 1\,,  \nonumber \\
&&
\Big[ \star \partial_t b(t, \vec{x}), c(t, \vec{y}) \Big]_+ =  -   {\frac  1 {\sqrt{g}}} \delta^3(x-y) \star 1\,,    \hspace{.5cm} 
\Big[ \star b(t, \vec{x}), \partial_t c(t, \vec{y}) \Big]_+ =   {\frac  1 {\sqrt{g}}} \delta^3(x-y) \star 1\, ,  \nonumber \\ &&
\end{eqnarray}
we can compute the action of the BRST charge on the fields
\begin{eqnarray}
\label{HEA}
\Big[Q^B, A\Big]_+ = dc\,, ~~~
\Big[Q^B, A^0\Big]= - \partial_t c\,, ~~
\Big[Q^B, c\Big]_+ = 0\,, ~~
\Big[Q^B, b\Big]_+ = { - \rho} \,  , ~~
\Big[Q^B, \rho\Big] =  0 .
\end{eqnarray}

We also define the local functional $\Phi(x)$ 
\beq
\Phi(x) = \int_{{\mathcal M}_3} \beta(x - y) d \star A(y) = \int \beta(x - y) \partial_i (g^{ij} A_j)(y) \sqrt{g}  d^3y .
\eeq{HD}
Note that $d \star A$ is a 3-form.
Finally, we can define the $R$-operator 
\begin{eqnarray}
  \label{HG}
R = \int_{{\mathcal M}_3} \Big( b \star A^0 + \partial_t b \star \Phi \Big) .
\end{eqnarray}
By computing the anticommutator between {$Q_0$ and $R$ we get 
\begin{eqnarray}
\label{HGA}
\Big[Q_0, R\Big]_+  = i \int_{{\mathcal M}_3} \Big(  - b \star \partial_t c + \partial_t b \star c 
+ i \rho \star A^0
+ i d \star E \wedge \Phi 
\Big) \equiv i S . 
\end{eqnarray}
}
This formula generalizes~\eqref{mass11} as it can be seen by defining $\Pi\equiv \star d\star E$. As in the previous
subsection, $\Pi$ and $\Phi$ are canonically conjugate when $\beta$ is the Green function of the Laplacian
\beq
\Delta \beta(x,y)= - {\frac  1 {\sqrt{g}}}\delta^3(x,y) .
\eeq{green1}
Then the counting operator $S$ becomes 
{
\beq
\Big[Q_0, R\Big]_+  = i \int_{{\mathcal M}_3} \Big(  - b \star \partial_t c + \partial_t b \star c 
+ i \rho \star A^0
+ i \star \Pi  \wedge \Phi \Big) . 
\eeq{HGC}
}

The formulas given here apply to open manifolds where equation~\eqref{green1} can be solved. In a closed 
manifold~\eqref{green1} is instead 
\beq 
\Delta \beta(x,y) = -  {\frac  1 {\sqrt{g}}} \delta^3(x,y) +{ \frac 1 {\rm{Vol}(\mathcal{M}_3)}} .
\eeq{lapl2}
{Furthermore, by integrating the first of eqs.~\eqref{HG'} on $\mathcal{M}_3$ we find that the total charge is a BRST-exact operator: }
\beq
\int_{\mathcal{M}_3} \star j^0= -\int_{\mathcal{M}_3}   \star  \partial_t \rho=
\left[Q^B, \int_{\mathcal{M}_3}   \star  \partial_t b \right] .
 \eeq{gausslaw}
This of course is simply the re-statement of the Gauss law.
 An analogous argument holds in perturbative gravity and it leads to the well known constraint that
 the matrix elements of the energy operator $H$ between physical (BRST-closed) states vanish for spatially closed 
 backgrounds, such as de Sitter spacetime.
 
 \section{Non-Abelian Gauge Theories}\label{nonab}

Let us begin by introducing some notation. We use the same geometric setting used for QED. In addition, the 1-form electric 
field $E = {E}^a t_a$ and the 2-form magnetic field $B = {B}^a t_a$ are vectors in the adjoint representation of a Lie 
algebra  with generators $t_a$.
We use the same notation for the electric potential (0-form) ${\phi}$ and the 1-form magnetic potential ${A}$. 
The fields  $E = {E}^a t_a, B={B}^a t_a$ are defined as follows 
\begin{eqnarray}
\label{NAA}
 {B} = d {A}  + \frac12 [{A}, {A}]\,, ~~~~~~~
 {E} =d_A A^0  - \partial_t {A} ,
\end{eqnarray}
where $d_A {A^0} = d {A^0} -  [{A}, {A^0}]$. 
${E}, {B}$ satisfy the Bianchi identities 
\begin{eqnarray}
\label{NAB}
d_A B =0\,, ~~~~~
d_A E + \partial_t B + [{A^0}, {B}]=0\, .
\end{eqnarray}
Here we used $d_A^2 {A^0} = [B, A^0]$ and the equations of motion
\begin{eqnarray}
\label{NAC}
d^\dagger_A E =  j^0 \,, ~~~~~~
d^\dagger_A B  - \partial_t E - [A^0, E]  = j ,
\end{eqnarray}
where $d^\dagger_A E  =\star d_A \star E$. $j^0$ is a 0-form which 
represents the charge density and  the 1-form $j$ is the current density.  Note that 
$d^\dagger_A  d^\dagger_A E = \star d_A^2 \star E = \star  [B, \star E]$. Acting with 
$d^\dagger_A$ on the second equation we have 
\begin{eqnarray}
\label{NAD}
d^\dagger_A d^\dagger_A B &=& \star [B, \star B] = d^\dagger_A j + d^\dagger_A \partial_t E + 
\star d_A \star[A^0, E] \nonumber \\
&=& d^\dagger_A j + \partial_t d^\dagger_A E - \star [\partial_t A, \star E] 
+ \star  [d_A A^0, \star E] + \star  [ A^0, d_A \star E] \nonumber \\
&=& d^\dagger_A j + \partial_t d^\dagger_A E - \star [(\partial_t A- d_A A^0, \star E] +   [ A^0, \star d_A \star E] \nonumber \\
& =& d^\dagger_A j + \partial_t j^0 + [A^0, j^0] =0 ,
\end{eqnarray}
where we used $\star [B, \star B] = \star [E, \star E]=0$. We can introduce the time-covariant derivative $\nabla_t j^0 = \partial_t j^0 + [A^0, j^0]$ such that the continuity 
equation becomes
\begin{eqnarray}
\label{NAE}
d^\dagger_A j + \nabla_t j^0 =0 .
\end{eqnarray}
The BRST transformations are 
\begin{eqnarray}
\label{NAF}
s\, A = d_A c = d c - [A, c] \,, ~~~~~~~
s\, A^0 = - \partial_t c - [A^0, c] 
\end{eqnarray}
and the fields $E, B$ transform covariantly 
\begin{eqnarray}
\label{NAG}
s\, B = [B, c]\,, ~~~~~
s\, E = - [E, c] .
\end{eqnarray}
By consistency the currents $j^0, j$ should transform as follows 
\begin{eqnarray}
\label{NAGA}
s j^0 = - [j^0, c] \,, ~~~~~
s j = [j, c] .
\end{eqnarray}
It is easy to verify that the equations of motion are indeed covariant under the BRST symmetry. 
Because of the nilpotency of the BRST charge, the ghost field $c$ transforms covariantly
\begin{eqnarray}
\label{NAGB}
s c = \frac12 [c,c]\   .
\end{eqnarray}

Now, we insert the potentials $A$ and $A^0$ into 
the equations of motion
\begin{eqnarray}
\label{NAH}
d^\dagger_A (d_A A^0 - \partial_t A) = j^0\,, ~~~~~~~
d^\dagger_A (d A + \frac12[A,A]) - \nabla_t (d_A A^0 - \partial_t A) = j  .
\end{eqnarray}
These equations are still gauge invariant, therefore we have to add the gauge fixing to the theory. As 
in the previous section, we choose the Lorentz gauge fixing { which adds to the Lagrangian a gauge fixing term}
\begin{eqnarray}
\label{NAI}
L_{GF}= \Tr \rho \star \left( \partial_t A^0 + d^\dagger A + \frac \xi 2 \rho \right).
\end{eqnarray}
Note that $d^\dagger_A A = d^\dagger A + \star[ A, \star A] =  d^\dagger A$. Therefore, we need to add the Nakanishi-Lautrup field 
$\rho$ and modify \eqref{NAH} as follows
\begin{eqnarray}
\label{NAHA}
d^\dagger_A (d_A A^0 - \partial_t A) - \partial_t \rho= j^0\,, ~~~~~~~
d^\dagger_A (d A + \frac12[A,A]) - \nabla_t (d_A A^0 - \partial_t A) + d \rho = j .
\end{eqnarray}

The BRST variation of 
the gauge fixing is given by 
\begin{eqnarray}
\label{NAL}
s \left( \partial_t A^0 + d^\dagger A \right) &=& \partial_t \left( - \partial_t c - [A^0, c] \right) - d^\dagger \left( d c - [A, c] \right) = 
- \partial^2 c - d^\dagger d c - \partial_t [A^0, c] + d^\dagger [A, c] \nonumber \\
&=& - \partial^2_t c + \Delta c - \partial_t [A^0, c] + d^\dagger [A, c]  ,
\end{eqnarray}
where the first two terms give the wave operator while the second two give the interaction terms between the ghost field and the 
potentials $A^0, A$. 

Let us write now the BRST charge separating the linear part from the higher order terms 
\begin{eqnarray}
\label{NAM}
Q_0 &=&  
 \int_{{\mathcal M}_3} {\rm Tr}\Big( dc \wedge \star E -   \star \partial_t c  \rho\Big) , \nonumber \\
Q_1 &=&  \int_{{\mathcal M}_3} {\rm Tr}\Big( - [A,c] \wedge \star E  -  \star  [A^0,c]  \rho  
+  \star  [b,c]  \partial_t c + \star \frac12    [c,c]   \partial_t b \Big)  .
\end{eqnarray}
Note that the BRST variation of the fields $b$ and $\rho$ are 
given by 
\begin{eqnarray}
\label{NAN}
s b = \rho + [b,c]\,, ~~~~
s \rho = [\rho, c] .
\end{eqnarray}
These transformations display the covariant properties of $\rho$ and $b$ fields, but they can be simplified by redefining $\rho$. 
Note that $d \star A$ is a 3-form. So we can finally define the $R$-operator 
\begin{eqnarray}
\label{NAP}
R = \int_{{\mathcal M}_3} {\rm Tr}\Big( b \star A^0 + \partial_t b \star \Phi[A] \Big) ,
\end{eqnarray}
where 
\begin{eqnarray}
\label{NAQ}
\Phi[A](x) = \int_{{\mathcal M}_3} \beta(x-y) d \star A(y) 
\end{eqnarray}
carries an index in the adjoint representation as $A$. 
{The (anti)commutation relations between the field component $E^a$, $A^a$, $A^{0\, a}$, $\rho^a$, $c^a$, $b^a$
have the same as for the Abelian case.}
\begin{eqnarray}
\label{NAQA}
&&
\Big[ \star \rho^{a}(t, \vec{x}), A^{0 b}(t, \vec{y}) \Big]=  - { \frac  {i k^{ab}}{\sqrt{g}}} \delta^3(x-y) \star 1\,,  \hspace{1cm} 
\Big[ \star E^a(t, \vec{x}), A^b(t, \vec{y}) \Big] =  - {\frac  {i k^{ab}}{\sqrt{g}}} \delta^3(x-y) \star 1\,,  \nonumber \\
&&
\Big[ \star \partial_t b^a(t, \vec{x}), c^b(t, \vec{y}) \Big]_+ =  -   {\frac  {k^{ab}}{\sqrt{g}}} \delta^3(x-y) \star 1\,,    \hspace{.5cm} 
\Big[ \star b^a(t, \vec{x}), \partial_t c^b(t, \vec{y}) \Big]_+ =   {\frac  {k^{ab}}{\sqrt{g}}} \delta^3(x-y) \star 1\,,  \nonumber \\ &&
\end{eqnarray}
where $k^{ab} = {\rm Tr}(t^a t^b)$ is the Killing-Cartan metric of the gauge group and $t^a$ are the generators of the 
gauge algebra in the adjoint representation.
 
Computing the anticommutator between $Q_0$ and $R$, we obtain the counting operator 
\begin{eqnarray}
\label{NAR}
S_{YM} = \int_{{\mathcal M}_3} {\rm Tr}
\Big(  - b \star \partial_t c + \partial_t b \star c + i \rho \star A^0 + i \star E_L \wedge A_L 
\Big) ,
\end{eqnarray}
which is equivalent to $S$ in \eqref{HGA} with an additional trace over adjoint indices.
It counts the states associated to the quartet mechanism $b,c, A^0$ and $A_L$ and their 
conjugate momenta $\partial_t c, \partial_t b, \rho, E_L$. Notice that the transverse part of $A_T$ and $E_T$ do not enter in the counting operator 
and therefore have vanishing $S$-charge. This allows us to further decompose the nonlinear part of $Q$, namely $Q_1$ 
into 
\begin{eqnarray}
\label{NAS}
Q_1 = Q_1^0 + Q^1_1 + Q^2_1 ,
\end{eqnarray}
where 
\begin{eqnarray}
\label{NAT}
Q^0_1 &=& 
  \int_{{\mathcal M}_3} {\rm Tr}\Big( - [A_T,c] \wedge \star E_L \Big) , \nonumber \\
Q^1_1 &=& 
  \int_{{\mathcal M}_3} {\rm Tr}\Big( - [A_L,c] \wedge \star E_L - [A_T,c] \wedge \star E_T  -  \star  [A^0,c]  \rho  
+  \star  [b,c]  \partial_t c + \star \frac12    [c,c]   \partial_t b \Big) , \nonumber \\
Q^2_1 &=& 
  \int_{{\mathcal M}_3} {\rm Tr}\Big( - [A_L, c] \wedge \star E_T  \Big) .
\end{eqnarray}
Fortunately, the first and the last term drop out since the transversal components of $E$ and of $A$ and the longitudinal one are orthogonal 
with respect to the background 3d-metric $g = g_{ij} dx^i dx^j$. 

\section{Perturbative Quantum Gravity}\label{pqg}

\subsection{Perturbative quantum gravity on Minkowski space}\label{mink}

We use the notations of ref~\cite{Jha:2022svf}, which is based on the historical papers~\cite{dirac2,ADM1,ADM2}. 
 In this subsection we consider pure gravity without matter or cosmological constant;  
 in the next one we introduce a cosmological constant and consider maximally symmetric spaces. 
Therefore, here we have only the following degrees of freedom 
\begin{eqnarray}
\label{MEA}
\{\gamma_{ij}, \pi^{ij}, N^i, N\} .
\end{eqnarray}
They are the 3d metric $\gamma_{ij}$ (six degrees of freedom), the conjugate momenta $\pi^{ij}$, and the Lagrange 
multipliers $N^i,\, N$, 
{in terms of which the four-dimensional metric $g_{\mu\nu}$ writes as ($\mu=0,i$ etc.)
\beq
g_{\mu\nu}=\begin{pmatrix} -N^2 + N_l N^l  & N_j \\ N_i & \gamma_{ij} \end{pmatrix}, \quad N_i =\gamma_{ij}N^j .
\eeq{4dmetric}
}

The ADM Hamiltonian is 
\begin{eqnarray}
\label{MEB}
H_{ADM} = \int_{x_0=t} d^3x \left\{ N\left[ - \sqrt{\gamma} \, {}^{(3)}\!R - \frac{1}{\sqrt{\gamma}}  \left(\frac{\pi^2}{2} - \pi^{ij} \pi_{ij} \right) \right]
+ N^i \left[ - 2 \nabla^j \pi_{ij}\right] \right\} .
\end{eqnarray}
Here ${}^{(3)}\!R$ is the 3d curvature and $\gamma=\det \gamma_{ij}$. In the presence of a cosmological constant 
$\Lambda$ the Hamiltonian constraint  $\phi$ acquires a new term proportional to  $\Lambda$. 
The Lagrange multipliers implement the three plus one constraints inside the brackets
\begin{eqnarray}
\label{MEC}
{\phi} &=& \left[ - \sqrt{\gamma}\, {}^{(3)}\!R - \frac{1}{\sqrt{\gamma}}  \left(\frac{\pi^2}{2} - \pi^{ij} \pi_{ij} \right) \right]\nonumber \\
{\phi}_i &=&  \left[ - 2 \nabla^j \pi_{ij}\right] .
\end{eqnarray}

These constraints satisfy the algebra of the commutation relations
\begin{eqnarray}
\label{MED}
\left[ \int d^3x \, \xi^i \phi_i, \int d^3x \, \chi^j \phi_j \right] &=& 
\int d^3x \, (\xi^k \nabla_k \chi^i - \chi^k \nabla_k \xi^i) \phi_i\,, \nonumber \\
\left[  \int d^3x \, \xi^i \phi_i, \int d^3x \, \tau \phi \right]&=& 
\int d^3x \, \xi^k \nabla_k \tau  \phi\,, \nonumber \\
\left[  \int d^3x \, \sigma \phi, \int d^3x \, \tau \phi \right] &=& 
\int d^3x \, \gamma^{kl} (\sigma \nabla_k \tau - \tau \nabla_k \sigma) \phi_k \,, 
\end{eqnarray}
where the commutators 
 are computed with the respect to dynamical variables $\gamma^{ij}, \pi_{ij}$, which obey the canonical commutator relations
 $[\gamma^{ij}, \pi_{kl}] = i \delta^i_{(k} \delta^j_{l)} \delta^3(x-y)$, and where 
$\xi^i, \chi^i, \sigma, \tau$ are local free parameters associated to the generators of the algebra. 
The next step is to convert them 
into ghost fields $c^i, c$ introducing also their conjugate variables $\Pi^c_i, \Pi^c$ such that 
\begin{eqnarray}
\label{MEF}
[c^i(x,t), \Pi^c_j(y,t)]_{+} = \delta_j^i \delta^3(x -y)\,, \hspace{1.5cm}
[c(x,t), \Pi^c(y,t)]_{+} = \delta^3(x - y)\,. 
\end{eqnarray}

We also introduce the conjugate fields to $N^i, N$, which we call $\rho_i, \rho$. 
To make their commutators canonical we introduce a background spacetime metric $\bar{g}_{\mu\nu}$ and the
harmonic gauge fixing term
\bea
\mathcal{L} &=&  \rho_\mu \phi^\mu + \frac \xi 2 \rho_\mu \rho^\mu + \mbox{ ghost Lagrangian} \nonumber \\
 \phi^\mu & = & \bar{D}_\nu \sqrt{-g} (g^{\mu\nu} -\bar{g}^{\mu\nu} ) , \quad \bar{D}_\mu \bar{g}_{\nu\rho}=0 .
\eea{hgf}
By expanding around the Minkowski background $\bar{g}_{\mu\nu}=\eta_{\mu\nu}$ we obtain $\bar{D}_\mu=\partial_\mu$
and  the canonical commutation relations
\begin{eqnarray}
\label{MEG}
[N^i(x,t), \rho_j(y,t)] = i \delta_j^i \delta^3(x - y)\,, \hspace{1.5cm}
[N(x,t), \rho(y,t)] = i \delta^3(x - y)\, .
\end{eqnarray}
The anticommutators of the $b$-ghost fields are
\begin{eqnarray}
\label{MEGA}
[\Pi^{b,i}(x,t), b_j(y,t)]_+ =  \delta_j^i \delta^3(x - y)\,, \hspace{1.5cm}
[\Pi^b(x,t), b(y,t)]_+ =  \delta^3(x - y)\, .
\end{eqnarray}
where $\Pi^{b, i}, \Pi^b$ are the conjugate momenta to $b_i, b$. 

Next, we  build the BRST charge 
\begin{eqnarray}
\label{MEH}
Q &=& \int_{x_0=t} d^3x \Big[ 
c^i \phi_i + c \phi 
+ \Pi^c_i (c^k \nabla_k c^i + \gamma^{ik} c \, \nabla_k c) + \Pi^c c^k \nabla_k c 
 \nonumber \\ 
&+&
\rho_i \Big(c^k \nabla_k N^i - N^k \nabla_k c^i + \gamma^{ik} (c \nabla_k N - N \nabla_k c) \Big) + 
\rho (c^k \nabla_k N - N^k \nabla_k c)  
\nonumber \\ 
&+& 
\Pi^{b,i} \Big(\rho_i + c^k \nabla_k b_i - b_k \gamma^{kj} \gamma_{il} \nabla_j c^l + (c \nabla_i b - b \nabla_i c) \Big) + 
\Pi^b (\rho + c^k \nabla_k b - b_k  \gamma^{kj} \nabla_j c)   
\Big]  . \nonumber \\ && 
\end{eqnarray}
The first line contains the ADM constraints and the ghost terms. Notice that the structure is dictated by the algebra 
\eqref{MED}. The second line guarantees that the Lagrange multipliers $N^i, N$ transform under the BRST symmetry. The total ghost number is one. It is not difficult, albeit tedious, to check the nilpotency of the BRST charge using 
the (anti)commutators (\ref{MEF},\ref{MEG}) and \eqref{MEGA}. Due to the structure of the BRST charge and thanks to the 
terms in the second line of~\eqref{MEH},
the Hamiltonian~\eqref{MEB} is BRST invariant $\{Q, H_{ADM}\} =0$ (up to a boundary term). 
The structure of the complete generators with 
the addition of $\rho^i$ and $\rho$ was originally found in~\cite{Castellani:1981us}. 

In order to apply the 
strategy outlined in the previous discussion, we need to split the BRST charge into a term quadratic in the fields and a remnant that begins at cubic order. 
To achieve this we have to expand the constraints $\phi$ and $\phi_i$ into a term linear in the field, $\phi_0, \phi_{0,i}$,
and the rest. 
Expanding the 3d metric $\gamma_{ij}$ around flat space, we can write the momentum constraint as 
$\phi_i =  - \partial^i \pi_{ij}$, where $\partial^i$ replaces the covariant 
derivative $\nabla^i$. On the other hand, the $\pi$-dependent terms in the $\phi$-constraint are quadratic so the 
only linear term we are left with is the linearization of the 
$\sqrt{\gamma}\, {}^{(3)}\!R$ term. 
It is also convenient to decompose the metric $\gamma_{ij}$ into 
a transverse-traceless part $\gamma^{TT}_{ij} $, a trace part  $\gamma^T$  and a longitudinal part $\gamma^{L}_{ij}$ 
as follows 
\begin{eqnarray}
\label{NEWC}
\gamma_{ij} &=& \gamma^{TT}_{ij} + \gamma^{T}_{ij} + \gamma^{L}_{ij}\, , ~~~~~~ \nonumber \\
&&\gamma^{TT}_{ii} = 0\,, ~~~~ \partial^i \gamma^{TT}_{ij} = 0 \,, ~~~~~  \partial^i \gamma^{T}_{ij} = 0 , \nonumber \\
&& \gamma^{L}_{ij} = \partial_i f_j +  \partial_j f_i ,
\end{eqnarray}
and similarly for the conjugated momenta \cite{ADM2,Kuchar:1970mu,Chowdhury:2021nxw}. 
Note that the terms of the decomposition \eqref{NEWC} are orthogonal, for instance 
\begin{eqnarray}
\label{NEWCA}
&&\int_{x_0=t} d^3x \sqrt{\gamma} \gamma^{TT, ij} \gamma^L_{ij} = - 2 \int_{x_0=t} d^3x \sqrt{\gamma} \partial_i\gamma^{TT, ij} f_j =0 ,\nonumber \\
&&\int_{x_0=t} d^3x \sqrt{\gamma} \gamma^{TT, ij} \gamma^T_{ij} = -\frac{1}{2} \int_{x_0=t} d^3x \sqrt{\gamma} \gamma^{TT, ij} 
\partial_i \partial_j \chi =0 , \end{eqnarray}
where we used integration by parts, the definitions \eqref{NEWC} and the auxiliary field $\chi$ satisfying the differential equation $(\Delta - 3)\chi = \gamma^T$. In addition, in the 
second line we used the tracelessness of $\gamma^{TT}$. We can decompose the conjugate momenta in the same manner. Commutation relations pair canonical coordinates and conjugate momenta according to their irreducible representations 
$[\gamma^{TT}_{ij}, \pi^{TT, kl}] = i \delta^k_{(i} \delta^l_{j)} \delta^3(x-y), \dots$  
Using this decomposition, we can write the linearized form of the constraints \eqref{MEC} as
\begin{eqnarray}
\label{NEWCB}
\phi^i_0 = - 2 \partial_k \pi^{L, ki} \,, ~~~~~~~
\phi_0 = - \partial_i \partial^i \gamma^T_{kk}  - \partial_i \partial_k \gamma^{L, ik} .
\end{eqnarray}
Using these definitions, we can construct the free BRST charge $Q_0$ as
\begin{eqnarray}
\label{MEHa}
Q_0 &=& \int_{x_0=t} d^3x \Big[ 
c^i  \phi^i_0  + c \phi_0 - \Pi^{b,i} \rho_i - { \Pi^{b} \rho}
\Big] ,
\end{eqnarray}
and finally, the $R$ operator as 
\begin{eqnarray}
\label{MEI}
R = \int_{x_0=t} d^3x \Big[ b_i N^i +  b N - \Pi^c_i \Phi^i -  \Pi^c \Phi \Big] .
\end{eqnarray}
It has negative ghost number and the two expressions $ \Phi^i, \Phi$ 
{must be defined as in previous cases to
have simple commutation relations with the linearized constraints. We can easily check that one such term is}
\begin{eqnarray}
\label{MEL}
\Phi^i =  \int_{x_0=t} d^3xy\beta^i_j(x-y) \partial_k \gamma^{L,jk}(y)\,, ~~~~~~
\end{eqnarray}
where $\beta^i_j(x-y)$ satisfies the Green function equation 
\begin{eqnarray}
\label{MEK}
\Delta \beta^i_j(x-y) + \partial^i \partial_k \beta^k_j = (\delta^i_k + \Delta^{-1} \partial^i \partial_k) \delta^3(x-y) .
\end{eqnarray}
The additional term in the right-hand side is needed for consistency. By contracting the equation with 
respect to $\partial_i$, we get 
$\Delta \partial_k \beta^k_j = \partial_j \delta^3(x-y)$, which is solved by $\Delta \beta^i_j = \delta^i_j \delta^3(x-y)$.

{The term $\Phi$ can be obtained from the harmonic gauge condition in analogy to the divergence $\partial_i A^i$ 
in Abelian gauge theories. We can also simply write down an educated guess for it; namely:}
\begin{eqnarray}
\label{MEKA}
\Phi =   \int_{x_0=t} d^3y \beta(x-y)  \gamma^{ij} \Gamma^0_{ij}(y)\,,
\end{eqnarray}
where the Christoffel symbols $\Gamma^\mu_{~\nu\rho}$ restricted to the 
hypersurface at $x_0 = t$ coincides with the  {\it extrinsic curvature tensor}  $\Gamma^0_{ij}  = K_{ij}$.
Therefore, in $\Phi$ we have the  trace of $K_{ij}$.\footnote{The gauge condition $K=0$ was used in \cite{Witten:2022xxp} to 
solve the Hamiltonian constraint by extremizing the renormalized volume of a Cauchy surface. See also~\cite{park}  
and the seminal work by York~\cite{york}.}
In order to compute the commutator between the BRST charge $Q_0$ and $R$, we have to compute the commutators
 between the linearized Hamiltonian constraint $\phi_0$
and the extrinsic curvature $K_{ij}$. The latter is related to the conjugate momenta $\pi_{ij} = \sqrt{\gamma} (K \gamma_{ij} - K_{ij})$.
So we get, at the linearized level,
\begin{eqnarray}
\label{MEKB}
\Big[  - \sqrt{\gamma} \, {}^{(3)}\!R(x), \frac{1}{2 \sqrt{\gamma}} (\gamma_{ij} \pi - 2 \pi_{ij})(y)  \Big] = i \partial_i \partial_j \delta^3(x -y)  .
\end{eqnarray}
Then, after computing the commutator between the BRST charge and $R$, 
we find that {if we choose $\beta(x-y)$ in $\Phi$ to be the Green function of the Laplacian, we reproduce}
 the counting operator $S$ for 
gravity. Note that for the diffeomorphism constraint $ (- 2  \partial^i \pi_{ij} )  \sim 0$ there is one partial derivatives in the constraint and one partial derivative in the 
gauge fixing in $\Phi^i$, they combine to provide the 3d Laplacian acting on $\beta^i_j(x-y)$, while in the case of the Hamiltonian constraint the two derivatives are coming 
from the variation of the 3d curvature. 

Computing the anticommutator between $Q_0$ and $R$ we get the following expression 
\begin{equation}
\label{NEWA}
[Q_0, R]_+ = i \int d^3x 
\Big(  \Pi^{b,i} b_i  + \Pi^{b} b  - \Pi^c_i  c^i - \Pi^c  c + i \rho_i N^i +  i \rho N +   i \phi_{0} \Phi + i \phi_{0,i} \Phi^i
\Big)=iS_{GR} ,
\end{equation}
where $\phi_0$ and $\phi_{0,i}$ are the linearized constraints \eqref{MEC}. 
Using the expressions of $\Phi$ and $\Phi^i$, we can expand the last two terms and rewrite $S_{GR}$ as follows 
\begin{eqnarray}
\label{NEWB}
S_{GR}= i \int d^3x 
\Big( \Pi^{b,i} b_i  + \Pi^{b} b  - \Pi^c_i  c^i - \Pi^c  c + i \rho_i N^i +  i \rho N  -  i \gamma^T \pi^T +  i \gamma^{ij L} \pi_{L,ij}
\Big) .
\end{eqnarray}
We note that the counting operator $S_{GR}$ assigns zero charge to transverse-traceless components of $\gamma^{TT}$ and 
$\pi^{TT}$, it assigns charge $+1$ to the trace part of $\pi^T$ and to the longitudinal mode of $\gamma^{L}_{ij}$, and -1 charge  to 
$\gamma^T$ (the trace of the conjugate momenta) and $\pi^L_{ij}$. As in the QED case, we have +1 charges for $c, c_i, \Pi^b, \Pi^{b,i}, N^i, N$ and -1 charges to $\rho_i, \rho, b, b_i, \Pi^c, \Pi^c_i$. 
Therefore, the quadratic terms of the complete BRST charge $Q$ have zero charges, and the rest have positive charges since the constraints $\phi, \phi_i$ are decomposed as 
\begin{eqnarray}
\label{NEWD}
\phi &=& \phi_0 + \pi^{TT ik} \pi^{TT}_{ik} + \frac14 \gamma^{TT}_{kl} \gamma^{TT, kl} + H_{k,k} 
\nonumber \\
\phi_i  &=& \phi_{0,i} +  \partial_i \gamma^{TT}_{kl} \pi^{TT,ik}  + H_{ik,k}
\end{eqnarray}
where the two functions $H_{ik,k}, H_{k,k}$ are quadratic functions of $\pi^{TT, ik}, \pi^T, \gamma^{TT}_{ij}, \gamma^L_{ij}$ 
(see \cite{Kuchar:1970mu,Chowdhury:2021nxw} for a more complete discussion).

\subsection{Perturbative quantum gravity on maximally symmetric spacetimes}\label{maxsym}

The introduction of a cosmological constant leaves the conjugate momenta unchanged, as it is independent of time 
derivative of the 3D metric, $\dot \gamma_{ij}$.  
The Hamiltonian constraint is modified while the momentum constraints are unchanged: 
\begin{eqnarray}
\label{MSSA}
{\phi}_\Lambda &=& \left[ - \sqrt{\gamma}\, ({}^{(3)}\!R -2 \Lambda)  - \frac{1}{\sqrt{\gamma}}  \left(\frac{\pi^2}{2} - \pi^{ij} \pi_{ij} \right) ,\right]\nonumber \\
{\phi}_i &=&  \left[ - 2 \nabla^j \pi_{ij}\right] .
\end{eqnarray}
In the case of a 3D spacetime with boundary we have to add the boundary Hamiltonian 
\begin{eqnarray}
\label{MSSB}
H_B = 2 \oint _{S_t} \Big[ N (\kappa - \kappa_0) + N_i \frac{\pi^{ij}}{\sqrt{\gamma}} r_j \Big] \sqrt{\sigma} d^2x ,
\end{eqnarray}
where $\kappa$ is the extrinsic curvature for the two-surface $S_t = \partial \Sigma_t$, where $\Sigma_t$ is a maximal 
Cauchy surface of the entire spacetime. 
Here $r_i$ is the normal vector of the two-surface and $\kappa_0$ (which is the extrinsic curvature of $\Sigma_t$ 
embedded into flat space) is introduced to  regularize the boundary term. 

In the case of  perturbation theory on maximally symmetric $AdS_4$ space we expand the metric as $g_{ij} = \gamma_{ij} + \kappa h_{ij}$ where 
$\kappa = \sqrt{8\pi G}$, while the background metric $\gamma_{ij}$ is diffeomorphic to the metric on a constant 
time slice of global $AdS_4$
\begin{eqnarray}
\label{MAA}
\gamma_{ij} dx^i dx^j = \frac{dr^2}{1+r^2} + r^2 d \Omega^2_{2} .
\end{eqnarray}
The background satisfies the following identities 
\beq
R_{ijkl} = \gamma_{il} \gamma_{jk}  - \gamma_{ik} \gamma_{jl}\,, ~~~
R_{ij} = (1-d) \gamma_{ij}\,, ~~~~R = - d(d-1)/2\,, ~~~. %
\nabla_i \nabla_j N = \gamma_{ij} N, \quad d=3,
\eeq{MAB}
 the cosmological constant is $\Lambda = -3 
 $ and we use the convention $l_{AdS} =1$. 
We use the same decomposition of the metric $h_{ij}$ and its conjugate momenta $\pi^{ij}$ in terms 
of $TT$, $T$ and $L$ terms we used in the previous subsection.
It is also convenient to redefine the momenta $\pi^{ij}$ as 
follows $\Pi^{ij} = \frac{\kappa}{\sqrt{\gamma}} \pi^{ij}$. We will use $\nabla_i$ to denote the covariant derivatives w.r.t. 
the background metric $\gamma_{ij}$. In terms of these definition, we can construct an expansion 
of the constrains,  following ref.~\cite{Chowdhury:2021nxw}
\begin{eqnarray}
\label{ExA}
\sqrt{g} \phi_i &=& \sqrt{\gamma} \phi^{(n)}_i + \mathcal{O}\Big( \kappa^{n-1} \Big)\,, ~~~~~~ n=0,1,2, \dots \nonumber \\
\sqrt{g} \phi_\Lambda &=& \sqrt{\gamma} \phi^{(n)}_\Lambda + \mathcal{O}\Big( \kappa^{n-1} \Big)\,, ~~~~~~ n=0,1,2, \dots 
\end{eqnarray}
where $ \phi^{(n)}_i, \phi^{(n)}_\Lambda$ contain all terms in the perturbative expansion up to $\mathcal{O}(\kappa^{n-2})$. 

Let us start with the momentum constraint. At zeroth order, it vanishes trivially on the AdS background 
$\phi^{(0)}_i(\gamma) =0$. 
The next two terms in the expansion are 
\begin{eqnarray}
\label{ExB}
\phi^{(1)}_i &=& - \frac{2}{\kappa} \gamma_{ij} \nabla_k \Pi^{L, jk} , \nonumber \\
\phi^{(2)}_i &=& - \frac{2}{\kappa} \gamma_{ij} \nabla_k \Pi^{L, jk}   + (\nabla_i h_{jk} - 2 \nabla_k h_{ij}) \Pi^{L, jk} - 2 h_{ij} \nabla_k \Pi^{L,jk}  . 
\end{eqnarray}
 Notice that according to these conventions, the third term is a subleading 
 $\mathcal{O}(\kappa)$ one. Inserting the expression for 
 $\phi^{(2)}_i$ in the decomposition of $h$ and $\Pi$ we get 
  \begin{eqnarray}
\label{ExC}
\phi^{(2)}_i - \phi^{(1)}_i  &=& \Big(2 (R_{klij} f^l - \nabla_k \nabla_j f_i) 
+ \nabla_i h^{TT}_{jk} - 2 \nabla_k h^{TT}_{ij}) \Big) \Big( \Pi^{ij}_T + \Pi^{ij}_{TT} \Big) .
\end{eqnarray}
We notice that the first-order piece of the momentum constraint $\phi^{(1)}_i$ replaces the flat space 
expression \eqref{NEWCB} and if we assign the charges as in the flat case, we see that it carries negative charge $-1$. 
Note that the background metric $\gamma_{ij}$ does not carry any charge since it is a classical background that commutes 
with all operators. In addition, we see that
in~\eqref{ExC} 
the second-order constraint $\phi^{(2)}_i$ carries the following charges: $0, +1, +2$. 
{The quadratic part of the BRST charge has always the general form~\eqref{COSE} while the interactions terms are
similar the Minkowski background case discussed in the previous subsection.}
Therefore, the first-order term $\phi^{(1)}_i$ enters the free BRST charge $Q_0$ and together with the 
charge of the ghost $c_i$ it gives it zero charge while all other terms in $Q^B$ carry {positive charge}.

We employ the same expansion for the Hamiltonian constraint. The zeroth order also vanishes due to the background metric. 
At first order we have 
\begin{eqnarray}
\label{ExD}
N \phi^{(1)}_{\Lambda} = - \frac{1}{2 \kappa} N \Big( \nabla^i \nabla^j h_{ij} - \nabla_i \nabla^i h) + 2 h \Big) = 
 - \frac{N}{2 \kappa}\nabla^i J_i[h^T] ,
\end{eqnarray}
where $J^i[h^T]$ is the ADM current and it depends upon the $h^T_{ij}$ part of the metric. The expression above can be written as a total derivative because of the lapse function $N$. However, this nice-looking expression allows us to deduce the total charge of this part of the Hamiltonian constraint. Notice that according to the 
charges assigned in the previous section, $N$ carries charge $+1$, while $h^T_{ij}$ carries unit negative charge. When  $N$ is 
stripped from~\eqref{ExD}, inserted into the BRST charge, and multiplied by the ghost fields $c^i, c$, this contribution to the free part of the BRST operator $Q_0$ has zero charge. The analysis of the second-order piece requires a little more effort. 
It turns out that the full expression for $N \phi^{(1)}_{\Lambda}$ 
is given by 
\begin{eqnarray}
\label{ExE}
N \phi^{(2)}_{\Lambda} - N \phi^{(1)}_{\Lambda}&=&  2 N  \Pi^{ij}_{TT} \Pi_{TT, ij} 
- \frac18 h_{TT}^{ij} (\Delta_N +2) h_{TT}^{ij} \nonumber \\
&+& \frac{1}{2} \nabla_i M^i + \frac{1}{4} \nabla_i L^i[h^{TT}] + \mathcal{O}(h^T) ,
\end{eqnarray}
where $M^i$ and $L^i[h^{TT}]$ are functions of $h^L, h^{TT}$ and $h^{TT}$ and $N$, respectively.
It can be checked that all terms have charges $0, +1$ or $2$ for $M^i$, while $L^i[h^{TT}]$ has zero charge. Therefore, 
the second-order terms in the Hamiltonian constraint has charges $0,1,2$ satisfying our criteria for constructing the 
intertwiner, {namely that the BRST charge decomposes into a quadratic term with zero $S_{GR}$ charge and 
higher-order terms with positive charges.}

As we have seen, both in the flat case and in the AdS space, the expansion of the constraint around the background 
allows us to verify that the constraints are function of the zero or positive-charged components of the tensors $\pi$ (or $\Pi$) 
and $\gamma$ (or $h$) 
according to our definition of charges. Nonetheless, for a more complete and background independent analysis we refer to 
\cite{Deser:1967zzb} where it is shown that the expansion of the constraints $\phi_i$ and $\phi$ into a linear part and a 
nonlinear part 
can be viewed as a method to fix the negative-charged components of the tensors $\pi$ and $\gamma$, namely $\pi^L, \gamma^T$, in terms 
of nonnegative-charge components $\pi^{TT}, h^{TT}, \pi^T, h^L$ . This translates into the decomposition of the constraints according to our charges. 
Schematically, we have 
\begin{eqnarray}
\label{ExEA}
\phi^i(\gamma, \pi) &=& \phi_{0}^i({\pi^L}) + \phi^i_{_{>0}}(\pi^{TT}, \gamma^{TT}, \pi^T, \gamma^L)\,,  \nonumber \\
\phi(\gamma, \pi) &=& \phi_0^i({\gamma^T}) + \phi_{_{>0}}(\pi^{TT}, \gamma^{TT}, \pi^T, \gamma^L)\,,  
\end{eqnarray}
where $\phi^i_{_{>0}}$ and $\phi_{_{>0}}$ are the nonlinear part of the constraint. A similar situation happens in the case of non-Abelian gauge theories, see eq.\eqref{NAT}, where the linear part of the constraint depends upon the 
negative-charge field $E_L$, but all the nonlinear terms 
have zero or positive charges. 
Finally, 
in \cite{Deser:1967zzb}, it is pointed out that the simplifications discussed here for AdS and flat backgrounds occur 
because they are
maximally symmetric spaces, satisfying $R_{ij} = \lambda g_{ij}$ with $\lambda$ constant. 
For the de Sitter case we refer to \cite{Chakraborty:2023los}, where a technique similar to the one we used here for $AdS_4$
 is employed.

\section{The Intertwiner}\label{inter}

\subsection{Construction of the intertwiner}\label{int1}

We can use the algebra found in the previous section to construct an operator that intertwines between the linearized BRST
charge $Q_0$ and the exact charge $Q^B$.
The key property we need is that in all cases described in the previous sections, namely QED, non-Abelian gauge theories and gravity, $Q^B$ decomposes into a finite  sum of terms with non negative $S$ charge:
\beq
Q^B=Q_0 +\sum_{0<n<N} Q_n , \quad N\in \mathbb{N}, \quad [S,Q_n]=nQ_n .
\eeq{mass17}

The operator defined below interpolates from $Q(0)=Q^B$ at $t=0$ to $Q(+\infty)=Q_0$ at $t=+\infty$, where we introduced
the definitions
\beq
Q(t)= Q_0 + Q_I(t), \quad Q_I(t)= \sum_{0<n<N}e^{-nt} Q_n.
\eeq{mass18}
Now the nilpotency of the BRST charge, $Q_B^2=0$, and the commutation relation $[S,Q_n]=nQ_n$ imply 
$\sum_{n+m=k} [Q_n,Q_m]_+=0$ for all 
$0\leq k <2N$ hence $Q(t)^2=0$ for all $t$ hence $Q_0 Q_I(t) + Q_I(t)Q_0 + Q_I(t)^2=0$.
Using the last identity we find
\bea
[[Q_I(t),R]_+ ,Q(t)] &=& Q_I(t) R (Q_0+Q_I(t)) + R Q_I(t) (Q_0+Q_I(t))  \nonumber \\ && -(Q_0+Q_I(t)) Q_I(t)R   -(Q_0+Q_I(t))RQ_I(t) \nonumber \\ &=&
Q_I(t)RQ_0 +Q_I(t)RQ_I(t) -RQ_0Q_I(t) +Q_I(t)Q_0R  \nonumber \\ && -Q_0RQ_I(t) -Q_I(t)RQ_I(t) \nonumber \\ &=&
[Q_I(t), [Q_0,R]_+]= i[Q_I(t),S]=-i \sum_{0<n<N}e^{-nt} n Q_n= i {dQ_I\over dt} .
\eea{mass14}
We define next an evolution operator $\Omega(t)$ by
\beq
{d\over dt} \Omega(t)= i \Omega(t) [Q_I(t),R]_+.
\eeq{mass19}
Thanks to~\eqref{mass14} we then find
\beq
{d\over dt} \Omega (t) Q(t) \Omega^{-1}(t) = \Omega (t)\left( i[[Q_I(t),R]_+ ,Q(t)]  + {dQ_I\over dt} \right) \Omega^{-1}(t)=0.
\eeq{mass20}
Integrating eq.~\eqref{mass20} from $t=0$ to $t=+\infty$ and using $\Omega(+\infty)=1$, $Q(+\infty)=Q_0$ we get
\beq
Q_0 -\Omega(0) Q^B \Omega^{-1}(0) =0,
\eeq{mass21}
which indeed says that $\Omega(0)$ is an intertwiner.

Notice that the key advantage of the construction is that the hard part of the work is to find an operator $R$ with the right
anticommutation relations with $Q_0$: $[Q_0,R]_+\propto S$ This is done in the linear BRST theory, which is universal.  All complexities due to nonlinear interaction terms are irrelevant because the existence of an intertwiner 
depends only on the $S$-charge of the interacting part of the BRST charge.

Notice also that the construction of $\Omega$ requires that the BRST charge is nilpotent. This is not true in anomalous 
gauge theories~\cite{marn} so our construction can only work for non-anomalous gauge theories. Needless to say, this
 is an important consistency check of our formalism.

\subsection{Some uses of the intertwiner}\label{int2}

An interesting application of our technique relates to the problem noticed in~\cite{randall}: even a formal evaluation of 
entanglement entropies for space-like separated parts of spacetime (a.k.a. subregions)  in gravity and gauge theory 
requires that the operators defined in one of them are independent of those defined in the other one. So local operators in
one subregion must commute with those in the other. The problem is that even in perturbation theory
no charged local gauge-invariant operator exists in gauge theories and no local physical 
operator {\em tout court} exists in gravity;
the dressing necessary to make them gauge invariant extends to the boundary of spacetime, so commutation of operators
based on space-like separated regions is no longer guaranteed. 

A solution proposed for gravity~\cite{randall} is to make 
gravity massive. This is possible in holographic theories defined on asymptotically Anti de Sitter (AdS) space with the 
transparent 
boundary conditions that are used to justify the island proposal~\cite{alm,mjt}, because the massless limit of 
massive AdS gravity is smooth~\cite{novvd}. In asymptotically flat spacetime this is no longer true;
instead, massive gravity in asymptotically flat spacetimes becomes
strongly interacting at a very low cutoff which, in terms of the Planck mass $M_{pl}$ and graviton mass $m$ 
reads~\cite{lpr} $\Lambda_3\propto  ({M_{Pl} m^2})^{1/3}$. No ultraviolet completion to an energy scale 
$E\sim M_{Pl}$ is known.  Finally, in de Sitter (dS) space there exists a gap $O(H)$  in the range of allowed masses for any 
unitary spin-2 field theory~\cite{hig}, so the graviton mass cannot be made parametrically smaller than the Hubble scale $H$.  

The operator $\Omega$ that we constructed in the previous sections gives another way to solve the problems associated
 with defining 
operators in a region (the island $I$ in figure~\ref{island}) which is disconnected from the asymptotic boundary of spacetime. 
We can define the physical algebra $\mathscr{A}$ of local operators in the region $I$ as all local operators in the 
cohomology of $Q_0$ and use $\mathscr{A}$ to perform all calculations needed for 
entanglement wedge reconstruction~\cite{pen}. So, by construction, the modular automorphism group, entropies etc. are all 
the ones defined for $\mathscr{A}$. The exact algebra of physical operators in $I$ is then by definition
$\mathscr{A}'=\Omega^\dagger \mathscr{A} \Omega$, affiliated operators are also defined by conjugation, and entropies in 
the exact theory are the same as those in $\mathscr{A}$. 

An advantage of our construction is that it depends only on the existence of a Cauchy surface $\Sigma$ but it is otherwise
independent of the form of the background spacetime, so it applies equally well to asymptotically AdS, Minkowski and dS
spaces. Another advantage is that it applies equally to gravity and gauge theories and it does not require them to be in
a spontaneously broken phase, as it is necessary for gravity to be if one chooses to apply the solution proposed 
in~\cite{randall}.

 \subsection{Partition function and counting of gauge invariant operators}

In the previous section we observed that we can construct observables in the exact algebra of physical states associated to the 
region $I$ of a Cauchy surface $\Sigma$ using the automorphism $\mathscr{A}'=\Omega^\dagger \mathscr{A} \Omega$.
A particularly simple yet nontrivial example is the counting of gauge invariant operators at a spacetime point $(x,t)$ in QED.
The intertwiner $\Omega$ constructed in the previous 
sections is also a {\em formal} map from the Hilbert space of the interacting theory to the free Hilbert space, for which we can 
easily counts the gauge invariant operators. 
Here it is convenient to use the Batalin-Vilkovisky approach~\cite{batvil} (see~\cite{wei} for an accessible introduction to the formalism). 
We refer to \cite{Grassi:2024kif} for a discussion between partition functions and Batalin-Vilkovisky approach. 
In that framework, one assigns a ghost number $G$, a fermion number $F$ and a scale dimension $S$ to each field and
antifield (i.e. source for the corresponding field)~\cite{batvil,wei}  as follows 
\begin{table}[htp]
\begin{center}
\begin{tabular}{|c|c|c|c|c|c|c|c|c|}
\hline
Fields & $c$ & $A_\mu$ & $\psi$ & $\psi^*$ & $A^\star_\mu$ & $c^\star$  \\
\hline
Scale dim & $0$ & $1$ & $3/2$ & $5/2$ & $3$ & $4$  \\
\hline
Ghost numb & $1$ & $0$ & $0$ & $-1$  & $-1$ & $-2$   \\
\hline
Fermion numb & $1$ & $0$ & $1$ & $0$ & $0$  & $0$ \\
\hline
\end{tabular}
\end{center}
\label{default}
\end{table}%

Here we have denoted with $\psi^\star, A^\star$ and $c^\star$ the sources (antifields) of the BRST transformations of $\psi, A_\mu$ and $c$. The gauge fixing discussed 
above is achieved by an anti-bracket canonical transformations introducing the ghost field $b$ and the Nakanishi-Lautrup field $\rho$. 
We proceed in two steps, first we give the formula for one-particle states (Hilbert states of physical degrees of freedom) and then we compute the multi-state formula for the Fock space. 

The formula for single-particle cohomology is \cite{Connes:2002ya,Connes:2004xy}
\begin{eqnarray}
\label{cohoA}
\mathbb{P}(t) = {\rm Tr}_{{\mathcal H}_0}((-1)^{F+1} t^S) = \frac{ 1 - 4 \, t + 8 \, t^{3/2} - 8 \,  t^{5/2} + 4 \, t^3 - t^4}{(1-t)^4} ,
\end{eqnarray}
where we assigned the fugacity $t$ as follows: $(-1)^{F+1} t^S$.

The coefficients of the polynomial in the numerator correspond to the dimensions of the Lorentz representation.
The denominator takes into account the derivatives $\partial_\mu$ with dimension $+1$.  We can rewrite the partition function as follows 
\begin{eqnarray}
\label{cohoAA}
\mathbb{P}(t) = \frac{(1-t^2)}{(1-t)^4} \left( 1 + t^2 - 4 t + \frac{8 t^{3/2}}{1+t}\right)  ,
\end{eqnarray}
where the factor $(1-t^2)$ represents the on-shell condition while the numerator $(1-t)^4$ represents the derivatives, 
the polynomial 
$1 + t^2 - 4 t$ represents the gauge field degrees of freedom, $8 t^{3/2}/(1+t)$ represents the fermionic degrees of freedom 
and finally the numerator 
$(1+t)$ takes into account the halving due to Dirac's equation. 
Taking the limit $t \rightarrow 1$ in the bracket we get $-2 +2$ which are the two gauge degrees 
of freedom and two fermionic degrees of freedom for a single particle, respectively.

Note that this partition function 
satisfies the duality (field/antifield duality) 
\begin{eqnarray}
\label{cohoAB}
\mathbb{P}(1/t) = - \mathbb{P}(t) .
\end{eqnarray}
The absence of a factor $t^\alpha$ in the duality transformation is due to $Q^BQ^B=0$, which is equivalent
to absence of a gauge anomaly~\cite{marn}.

Let us expand $\mathbb{P}(t)$ in a Taylor series in $t$ and look at  the few first terms 
\begin{eqnarray}
\label{cohoB}
\mathbb{P}(t) = 1 + 8 t^{3/2} - 6 t^2 + 24 t^{5/2} - 16 t^3 + 48 t^{7/2} - 30 t^4 + O(t^4)  ,
\end{eqnarray}
where each single term has  the following interpretation 
\begin{eqnarray}
\label{cohoC}
 8 t^{3/2}  &\leftrightarrow& \psi, \bar\psi , \nonumber \\
 - 6 t^2  &\leftrightarrow& F_{\mu\nu} ,  \nonumber \\
 + 24 t^{5/2} &\leftrightarrow& \partial_\mu\psi, \partial_\mu\bar\psi ~ ~{\rm modulo~the~EOM:} ~~\gamma^\mu \partial_\mu \psi = \gamma^\mu \partial_\mu \bar\psi =0 , \nonumber \\
 - 16 t^3 &\leftrightarrow& \partial_\rho F_{\mu\nu} ~~ {\rm modulo~the~EOM~and~B.I.:} ~~
 \partial^\mu F_{\mu\nu} = \partial_{[\rho} F_{\mu\nu]} =0 ,  \nonumber \\
 + 48 t^{7/2} &\leftrightarrow& \partial_\mu\partial_\nu\psi, \partial_\mu\partial_\nu\bar\psi ~ ~{\rm modulo~the~EOM:} ~~\gamma^\mu \partial_\mu \partial_\nu\psi = \gamma^\mu \partial_\mu \partial_\nu\bar\psi =0  , \nonumber \\
 - 30 t^4 &\leftrightarrow& \partial_\rho \partial_\sigma F_{\mu\nu} ~~ {\rm modulo~the~EOM~and~B.I.:} ~~
\partial_\sigma \partial^\mu F_{\mu\nu} = \partial_{(\sigma}\partial_{[\rho)} F_{\mu\nu]} =0 . 
\end{eqnarray}
This is the correct BRST cohomology modulo the equations of motion and the Bianchi identities. 

To construct the multi-particle cohomology, we use the {\em plethystic polynomials}
\cite{Benvenuti:2006qr,Feng:2007ur}. First we rewrite the partition function \eqref{cohoA} 
in terms of a series 
\begin{eqnarray}
\label{cohoE}
\mathbb{P}(t) = \sum_{k=0}^\infty N_{k/2} t^{k/2} ,
\end{eqnarray}
where $N_{k/2}$ are integers with $N_0=1$. 
Then, we construct the multi-state partition function using the 
plethystic polynomial as follows 
\begin{eqnarray}
\label{cohoF}
PE[\mathbb{P}(t) x] =  \prod_{k=0}^\infty (1 - t^{k/2} x)^{N_{k/2}} .
\end{eqnarray}
The parameter $x$ counts the number of fields. Expanding in powers of $x$, we have 
the number of fields. Additionally, the powers of $t$ take into account in details the various gauge invariant operator. 
In the case of QED there are no multitrace operators since we do not have color factors. 

\section{Additional Remarks}\label{other}
\subsection{Other definitions of physical operators}\label{otherdef}
The reason why local physical operators can be defined in massive gravity is not special to gravity alone and works 
also for massive
gauge theories: both possess extra perturbative degrees of freedom. They are the St\"uckelberg (a.k.a. Goldstone) modes
corresponding to the longitudinal degrees of the massive gauge particle. They allow to write an explicit mass term in a 
gauge invariant way. In linearized gravity we expand the metric as $g_{\mu\nu}= \bar{g}_{\mu\nu} +M_{Pl}^{-1} h_{\mu\nu}$ around a
background $\bar{g}_{\mu\nu}$ and replace the fluctuation of the graviton, $h_{\mu\nu}$, with
\beq
h_{\mu\nu}\rightarrow \hat{h}_{\mu\nu}= h_{\mu\nu} + D_\mu \zeta_\nu + D_\nu \zeta_\mu .
\eeq{other1}
The nonlinear completion of the Pauli-Fierz mass term $\frac 1 2 m^2 (\hat{h}_{\mu\nu}\hat{h}^{\mu\nu}- \hat{h} \hat{h} )$
exists (see e.g.~\cite{drgt}) but its explicit form is not necessary here.
For non-Abelian gauge theories with fundamental representation indices ${}^a$ and dual indices ${}_b$ the St\"uckelberg modes are matrices $G^a_b$ and the gauge-invariant mass term is
\beq
{1\over 2} m^2 \Tr G^{-1} D_\mu G G^{-1} D^\mu G, 
\eeq{other2}
where $D_\mu$ is the covariant derivative. 

The mass term is gauge invariant because $G^{-1} D_\mu G$ itself is invariant under the gauge transformation acting as
$G\rightarrow G'=g^{-1}G$, $A_\mu\rightarrow A'_\mu=g^{-1} \partial_\mu g + g^{-1} A_\mu g$. So $G$ is the dressing
field that makes the gauge potential itself gauge invariant. The dressing that makes the charged matter fields $\psi^{a_1,..a_n}_{b_1,..b_m}$ gauge invariant is
 \beq
 \psi^{a_1,..a_n}_{b_1,..b_m}\rightarrow \hat{\psi}^{a_1,..a_n}_{b_1,..b_m}= G^{a_1}_{p_1}... G^{a_n}_{p_n} 
 (G^{-1})^{q_1}_{b_1}.... (G^{-1})^{q_m}_{b_m}\psi^{p_1...p_n}_{q_1,...q_m} .
 \eeq{other3}
 
 A similar but slightly more involved dressing procedure can be used for gravity. The nonlinear St\"uckelberg
 fields are coordinate transformations on the spacetime manifold $M$ that in a local neighborhood are explicitly
 given by $\phi^\mu(x) =x^\mu + M_{Pl}^{-1}\zeta^\mu(x)$ so they are invertible in perturbation theory where 
 $|\zeta^\mu|/M_{Pl} \ll 1$. Under a coordinate change $\psi : \;  M\rightarrow M$ a tensor field $T$ pulls back to
 $T_*^\psi$ so the dressed field defined as the pullback under $\phi : \;  M\rightarrow M$, $T_*^\phi$, is diffeomorphism 
 invariant if $T_*^{\phi_\psi \circ \psi}=T_*^\phi$. This defines the transformation law of the St\"uckelberg:
 $\phi_\psi \circ \psi = \phi$.\footnote{This is essentially the same construction first described in ref.~\cite{komar} and 
 periodically resurfacing in the literature in various forms~\cite{ah-g,hg,j-r}. To the best of our 
 knowledge, the non-Abelian broken gauge theory dressing~\eqref{other3} was first explicitly proposed  in~\cite{br}. 
 It is a consequence of the general method presented in~\cite{weinberg}.}
 
 The dressing procedure outlined here requires a gauge theory to be in the broken symmetry (Higgs) phase everywhere
  in spacetime so it runs into trouble when broken and unbroken phases coexist, or when a theory is simply in an unbroken 
  phase. In gravity this requirement is the same as demanding the existence of a clock (and yardsticks) at each point in
  spacetime. This is not necessary to define an algebra of local operators for the spacetime region accessible to an observer
  living on a time-like curve $\gamma$. Thanks to timelike-tube theorem~\cite{borch,witt-reg,wit-tube} the local algebra
  of operators in the time-like envelope $\mathcal{E}(\gamma)$ of $\gamma$ coincides with the algebra of time-smeared 
  operators in $\gamma$, so only operators in $\gamma$ must be made diffeomorphism invariant. All that is needed is a 
  one-dimensional St\"uckelberg field defined only on $\gamma$. For gravity this is the ``clock'' defined in~\cite{clpw}. 
  In gauge theory one can equivalently define a local St\"uckelberg field $\phi$ on $\gamma$ only. 
  When $\gamma$ is described by coordinates $x^\mu(\tau)$ its action is
  \beq
  S=\int d\tau \frac 1 2 \Big( {d\phi \over d\tau} -A_\tau(\tau)\Big)^2 , \quad A_\tau={dx^\mu\over d\tau} A_\mu(x(\tau)).
 \eeq{other4}
 This is a less radical change to a gauge theory than demanding the existence of a St\"uckelberg field everywhere in 
 spacetime, but it is still a dynamical property that does not hold for a generic gauge theory on an arbitrary background.
 Even if the clock is a dynamical part of a physical system, as in~\cite{pen-clock,pap-clock}, its existence is confined to a
 ``code'' subspace, as it is typical of relational observables.
 
{ \subsection{Closed maximal Cauchy surfaces}\label{closed}
 When the Cauchy surface used to define the charges $Q_0,Q^B,R$ is closed our construction must be modified. To 
 understand the necessary changes we will look at the Abelian gauge theory case, which is the technically the simplest but 
 already shows all the relevant features. First of all we notice that eq.~\eqref{lapl2} and the definition of the operator
  $\Phi(x)$~\eqref{HD} require
  \beq
  \int_{\mathcal{M}_3} d \star A =0 .
  \eeq{clos1}
  This equation in turns implies that $\int_{\mathcal{M}_3}  \star \partial_t b$, the operator canonically conjugate to the zero 
  mode of $c$,
  does not appear in $R$ and hence in the counting operator $S$ in~\eqref{HGA}. The operator $\Omega$ 
  constructed in~\ref{int1} in this case intertwines between BRST charges that do not contain the $c$-zero mode 
  \beq
  c_0={1\over {\rm{Vol}(\mathcal{M}_3)}}\int_{\mathcal{M}_3}  \star c. 
  \eeq{clos2}
  The free BRST charge $Q_0$ is independent of $c_0$ because 
  $\int_{\mathcal{M}_3} d\star E =0$ on a closed Cauchy surface. On the other hand $Q_1$ depends on it because
  \beq
  Q_1 = c_0 \int_{\mathcal{M}_3} \star j^0 + Q'_1 , \quad Q'_1 \mbox{ independent of } c_0.
  \eeq{clos3}
  So the charges $Q^B$ and $\Omega^\dagger Q_0 \Omega$ agree only on the subspace of states $\Psi$ obeying 
  $\int_{\mathcal{M}_3} \star j^0 \Psi =0$. This is a stronger condition than~\eqref{gausslaw}. For gravity on de Sitter space
  space the constraint is $H\Psi=0$.
   }
 \subsection{States}\label{states}
 The intertwiner $\Omega(0)$ defined in eq.~\eqref{mass21} can be used to construct states that formally satisfy the 
 physical state condition. On any spacetime that possesses a nonzero timelike Killing vector defined in an 
 open neighborhood of the Cauchy surface $\mathcal{M}_3$ we can define creation and annihilation operators for matter
 states, for the gauge fields (in QED and Yang-Mills theories) or the graviton (in gravity), as well as for the ghost fields.
 Calling them collectively $a$ and $a^\dagger$ we can then define the standard Fock vacuum by $a|0\rangle =0$.
 In the presence of zero modes for the $b$- and $c$-ghosts, the vacuum can carry a ghost number, which can be used to
give different definitions of physical states (see e.g. appendix B of~\cite{clpw}). 
This is true irrespectively of the BRST charge used to
define physical states --which are vectors in the cohomology of the BRST operator. 

Let's start by constructing states in the cohomology of $Q_0$ at fixed ghost number. A large class of
such states is extremely simple to construct. These are {\em all} the Fock-space
states built out of the vacuum by applying creation operators of matter fields and transverse gauge fields (or transverse and
traceless gravitons). If the Fock vacuum we choose is annihilated by all the $b$-ghost zero modes, the cohomology defines
 states invariant under the linearized gauge transformations (or diffeomorphisms), while if the Fock vacuum is annihilated by
 all the $c$-ghost zero modes, it belongs to the space of co-invariant states. This construction is reviewed for instance
  in appendix B of~\cite{clpw}. Whichever choice one makes, the point is that a large vector space $V$ can
  be constructed, which belongs to the cohomology of $Q_0$. Let's call $\Psi_0$ a generic nonzero state in $V$. 
  Then the state $\Psi=\Omega^\dagger\Psi_0$ obeys
  \beq
  Q^B\Psi=Q^B\Omega^\dagger \Psi_0=\Omega^\dagger Q_0 \Psi_0=0,
  \eeq{stat1}
  so  $\Psi$ is $Q^B$ closed. If it were exact $\Psi=Q^B \Xi$, then 
  $\Psi_0=\Omega Q^B \Xi =  Q_0  \Omega \Xi$ would be exact, hence zero in the $Q_0$ cohomology. 
  
  So far all has been simple, what is less obvious and left to future investigations is which of these formal manipulations
  are justified. To conclude on a downbeat, let's list some of the problems that may be encountered in trying to make our 
  construction precise.
  \begin{enumerate}
  \item The operator $\Omega$ may not exist as a unitary operator on the Fock space. This is the fate of the  M\o{}ller operator
  in perturbation theory~\cite{haag,rs}. 
  
  Note that $\Omega$, better the infinite-volume limit of an IR-regulated operator,
   can still define an automorphism of the algebra
   of local observables. The idea is to note that the charges $Q^B,Q_0,R$ are integrals on a Cauchy surface of 
   (possibly nonlocal) 
  currents. One can define charges restricted to integration over a finite volume $V$, as $Q^B_{V}=\int_V d^3x j^0_{BRST}$ etc.
  and use them to define an IR-regulated operator $\Omega_V$. The limit
  \beq
  O'(x)=\lim_{V\rightarrow \infty} \Omega_V^\dagger O(x) \Omega_V
  \eeq{stat2}
  may exist even when the limit $\lim_{V\rightarrow \infty} \Omega_V$ does not.
  \item
  We need to define a norm for $\Psi$. One possibility is to define it in term of the norm $(A,B)_0$ on the Fock space  of the $a,b,c$ free fields as $(\Psi,\Phi)\equiv (\Omega \Psi, \Omega \Phi)_0=(\Psi,\Phi)_0$, but this makes sense only if 
 $\Omega$ exists as an operator of the Fock space --which is a tall order in any field theory. 
  \item 
 A related problem is that even if $\Omega$ exists, states of the form  $\Psi=\Omega^\dagger \Psi_0$ may not satisfy other
 criteria for being physical, such as having finite energy in gauge theory or finite ADM energy in gravity.
 \end{enumerate}
 \subsection*{Acknowledgements} 
M.P. is supported in part by NSF grant PHY-2210349. During part of this work MP was also supported by the 
Leverhulme Trust through a Leverhulme Visiting 
Professorship at Imperial College, London.

\appendix
\section{Constraints, BRST Symmetry and the $R$ Operator}\label{app}
In this appendix we give a general overview of the constructions used in the paper. We consider a set of 
constraints $\phi_I$ with $I=1, \dots, N$ which form the first class constraint algebra formed by computing 
 the commutator
\begin{equation}
\label{COSA}
[\phi_I, \phi_J] = f_{IJ}^{~~K} \phi_K ,
\end{equation}
where $f_{IJ}^{~~K}$ are the structure constants which satisfy 
the Jacobi identities.  We introduce the Lagrangian multipliers $A^K_0$, a set of anticommuting fields $c^K, b_K$, and 
their conjugate momenta $\rho_K$ and $\Pi^c_K, \Pi^{b,K}$ such that 
\begin{eqnarray}
\label{COSB}
&&\Big[\rho_K(x,t), A^L_0(y,t)\Big] = -  i \delta^L_K \delta^3(x-y)\,, ~~~ \nonumber \\
&&\Big[\Pi^c_K(x,t), c^L(y,t)\Big]_+ =  \delta^L_K  \delta^3(x-y)\,, ~~~
\Big[\Pi^{b,K}(x,t), b_L(y,t)\Big]_+ =  \delta_L^K  \delta^3(x-y)\,  .  ~~
\end{eqnarray}
Let us define the BRST charge 
\begin{eqnarray}
\label{COSC}
Q = \int_{{\mathcal M}_3}\Big[  c^K \phi_K + \frac12 f_{KL}^{~~M} \Pi^c_M  c^K c^L +   f_{IK}^{~~M} \rho_M c^I  A^K_0 + \Pi^{b,I} 
\left(\rho_I  +  f_{IK}^{~~M}  c^K b_M  \right)\Big] \, ,
\end{eqnarray}
which, acting on the fields, gives 
\begin{eqnarray}
\label{COSD}
&&\Big[Q, \phi_K\Big] = f_{LK}^{~~\,M} c^L \phi_M\,, ~~~~~ \Big[Q, c^K\Big]_+ = \frac12 f_{IJ}^{~~\,K} c^I c^J\,, ~~~~~~~~\nonumber \\
&&\Big[Q, b_K \Big]_+ = \rho_K + f_{KL}^{~~\,M} c^L b_M  \,, ~~~~~~~~\Big[Q, A^K_0\Big] =\Pi^{b,K} +  f_{IJ}^{~~\,K} c^{I} A^J_0\,, ~~~~~~\nonumber \\
&&\Big[Q, \Pi^c_K \Big]_+ = \phi_K+ f_{KL}^{~~\,M} c^L \Pi^c_M \,, ~~~~~~~~ \Big[Q, \Pi^{b,K} \Big]_+ =  f_{IJ}^{~~\,K} c^I \Pi^{b,J}  ,~~~~~~~~ \nonumber \\
&&\Big[Q, \rho_{b,K} \Big]_+ =  f_{IJ}^{~~\,K} c^I \Pi^{b,J} 
 .
\end{eqnarray}
 Now, we decompose the BRST charge 
 as 
 $$Q = Q_0 + Q_1   ,  $$
 where $Q_0$ is the linear part of the symmetry generated by the constraints $\phi_K =  \phi_{1,K} + \dots $ and $Q_1$ are 
 the higher order terms. Schematically, 
 we have 
 \begin{eqnarray}
\label{COSE}
Q_0 = \int_{{\mathcal M}_3}\Big[  c^K \phi_{1,K}  + \Pi^{b,I}  \rho_I  \Big] ,
\end{eqnarray}
which is nilpotent since $\phi_{1,K}$ form an Abelian algebra.

 The $R$ operator is defined as follows 
 \begin{eqnarray}
\label{RopA}
R =  \int_{{\mathcal M}_3}  \left( A_0^I b_I + \Pi^c_K \Phi^K \right) ,
\end{eqnarray}
where the field $\Phi^K$ has the transformation property 
\begin{eqnarray}
\label{RopB}
\Big[Q_0, \Phi^K \Big] = c^K . 
\end{eqnarray}
This is a ``naked" ghost field, namely {the result} of a BRST transformation which varies a field into a ghost 
without any differential/algebraic operator acting on $c^K$. This is usually classifed as 
a ``topological" symmetry which allows us to remove $\Phi^K$ and $c^K$ from the cohomology using the BRST doublet theorem. 
Suppose that we have chosen a gauge fixing function 
$\Psi^k$ such that $[ \phi_{1,K}, \Psi^L] = i \Delta \delta^3(x-y)$ and we have chosen $\beta(x-y)$ to be 
the Green function of $\Delta$, then
we can 
write $\Phi^I$ as follows 
\begin{eqnarray}
\label{RopC}
\Phi^K =  \int_{{\mathcal M}_3} \beta(x-y) \Psi^K(y) .
\end{eqnarray}
Then we have 
\begin{eqnarray}
\label{RopD}
\Big[Q_0, R\Big]_+ = i  \int_{{\mathcal M}_3}  \Big( -\Pi^c_K c^K + \Pi^{b,K} b_K + i A^K_0 \rho_K + i \phi_{1,K} \Phi^K \Big) ,
\end{eqnarray}
which is the counting operator discussed in the text. 
 


\end{document}